\documentclass[twocolumn,pra,aps]{revtex4-1}

\usepackage{color}
\usepackage{epsfig}
\usepackage{latexsym}
\usepackage{amssymb}
\usepackage{amsmath}

\usepackage{algorithm}
\usepackage{algorithmic}
\usepackage{wrapfig}
\usepackage[apple mac]{inputenc}
\usepackage[english]{babel}
\usepackage{times}
\usepackage{latexsym}
\usepackage{fancyhdr}
\usepackage{verbatim}
\usepackage{tabularx}
\usepackage{epsfig}
\usepackage{amsmath}
\usepackage{amssymb}
\usepackage{graphicx}
\usepackage{wasysym}

\usepackage{yfonts}

\newcommand{\qed}{\hspace*{\fill}$\square$}

\newcommand{\be}{\begin{equation}}
\newcommand{\ee}{\end{equation}}

 \newcommand{\half}{\frac 1 2}

\begin{document}

\title{A General Relativistic Pendulum: Isochronous vs. Geodesic Motion}
\author{Miguel A. Martin-Delgado}
\affiliation{Departamento de F\'{\i}sica Te\'orica, Universidad Complutense, 28040 Madrid, Spain.\\
CCS-Center for Computational Simulation, Campus de Montegancedo UPM, 28660 Boadilla del Monte, Madrid, Spain.}

\begin{abstract} 
General Relativity (GR) is shown to be  a complete theory with respect to the isochrony of the pendulum. This guarantees that time can be measured with a mechanical clock within the theory itself as a matter of principle.
The proper and coordinate oscillation frequencies of a simple pendulum are computed as a function of 3 coordinate parameters: the distance of the fulcrum to the gravity source, its length and the Schwarzschild radius. Novel GR effects appear in the Schwarzschild coordinates as divergences,   zeroes and other instabilities  of the oscillation frequencies that are absent in Newtonian gravity.
The anomalies in the proper and coordinate frequencies occur under extreme conditions: either the fulcrum or the pendulum mass or both remain inside the Event Horizon.

\end{abstract}

\maketitle

\tableofcontents

\section{Introduction}
\label{sec:intro}

We pose the question of how to measure time in General Relativity (GR) using mechanical clocks that abide by the laws of the theory.  In GR, time is relative to the observer whether it is proper time, coordinate time or else but the device to measure it is never specified. This is an ontological question about General Relativity. We could just say that we use atomic or optical clocks. However, this would put the device outside GR theory since those clocks abide by the laws of quantum mechanics. Thus, in a strict sense, GR would not be a complete theory if we cannot measure time with mechanical clocks fulfilling the laws of General Relativity.

Notice that the question of measuring time in General Relativity \cite{Einstein_relativity}, or in any other classical theory, by means of a mechanical device (clock) obeying the same theory's laws has a different nature from the notion of gravitational or relativistic time dilation: these ones refer to comparisons of clocks by different observers measuring different proper times. Thus, the measurement of proper times is prior to their comparison.

True as it is, once we have a device that is isochronous it can work as a clock no matter its origin,
whether it is classical or quantum.  If GR were not isochronous it would pose the theory in an awkward position.
As Einstein once stated: time is what you measure with a clock \cite{Gilder} (much as space is what you measure with a ruler). These are operational definitions specially well-suited for metrology.

After answering this question in the affirmative by studying the compatibility of isochrony with geodesic motion, we address the issue of the accuracy of mechanical clocks based on the simple pendulum.
We find that whenever an anomaly occurs in the functioning of the mechanical clock, this happens when either the support point of the pendulum or its mass or both are inside the Event Horizon (EH).

It is shown that the geodesic equations of motion for a simple pendulum in the gravitational field
described by the Schwarzschild metric admits isochronous motion in the regime of small amplitude oscillations.  This allows us to find the expression of the proper and coordinate oscillation frequencies in terms of coordinate parameters such as length of the pendulum,  the distance of its support point to the mass of the gravitation field and the corresponding Schwarzschild radius.  We provide several consistency checks like reproducing the period of a simple pendulum in the Newtonian gravity and the celebrated Huygens formula for the pendulum in a constant gravity field.

\section{Isochronous and Geodesic Motion}
\label{sec:Isochronous}

On general grounds,  to answer the question of whether we can construct a mechanical clock
obeying the laws of General Relativity is equivalent to  searching for an isochronous motion that is compatible with being also geodesic.  Isochrony was  the crucial property that Galileo discovered in the movement of the pendulum that allowed the measurement of time with an accuracy of 1 $s$ for the first time.  Before that,  the second was defined by a division of the day measured by the Earth's rotation.  With isochrony it is possible to measure time by counting the number of periods of a pendulum for a given event.  Isochrony means that the period is independent of the initial amplitude of the pendulum,  which is true for small oscillations.  

One possibility is to search from scratch an isochronous motion that is compatible with being also geodesic.  
This amounts to the existence of an harmonic motion of frequency $\omega$ compatible with GR. Let $Q=Q(x^{\mu})$ be the generalized coordinate of a degree of freedom of a particle of mass $m$ obeying the harmonic oscillator equation:
\begin{equation}
\frac{d^2 Q(x^{\mu})}{d\tau^2} + \omega^2 Q(x^{\mu}) =0,
\label{harmonic1}
\end{equation}
where $x^{\mu}$, $\mu=0,1,2,3$ are the space-time coordinates specifying the particle's motion in a certain gravity field. The derivative is with respect to proper time $\tau$ and the period of oscillation is $T=\frac{2\pi}{\omega}$ defines the mechanical clock. The particle's coordinates must satisfy the condition of geodesic motion in the presence of gravitation
\begin{equation}
\frac{d^2 x^{\lambda}}{d\tau^2} + \Gamma^{\lambda}_{\mu \nu} \frac{d x^{\mu}}{d\tau}  \frac{d x^{\nu}}{d\tau} = 0.
\label{geodesics}
\end{equation}
Alternatively,  the search of $Q(x^{\mu})$ can be recast in the form
\begin{equation}
 \frac{\partial^2 Q}{\partial x^{\mu}x^{\nu}} \dot{x}^{\mu} \dot{x}^{\nu} + \frac{\partial Q}{\partial x^{\mu}} \ddot{ x}^{\mu} +\omega^2 Q(x^{\mu}) =0,
 \label{harmonic2}
\end{equation}
This is a second order  linear partial differential equation in $Q(x^{\mu})$ whose coefficients obey the geodesics equations,  which are non-linear in turn.  Thus, solving for \eqref{harmonic2} and \eqref{geodesics} may become a daunting task.

\section{The Simple Pendulum in General Relativity}
\label{sec:Pendulum}

Another alternative to seeking isochronous and geodesic motion is to start with examples of geodesic motion associated with well-known solutions of the Einstein field equations of GR and try to search for isochronous motions among them.  For sure, there are well-known instances of this combined motion like circular orbits of photons (the Photon Sphere in the range $r\geq  \frac{3 GM}{c^2}$) or massive particles orbiting around a gravitating mass in the range $\frac{6GM}{c^2}\leq r < \infty$ \cite{Misner,  Carroll}.    However,  those orbits are unstable for photons while they are stable for massive particles.  Nevertheless,  the latter stability is not enough to provide a controllable mechanism to construct a clock like the pendulum does.

\begin{figure}[t]
  \includegraphics[width=0.5\textwidth]{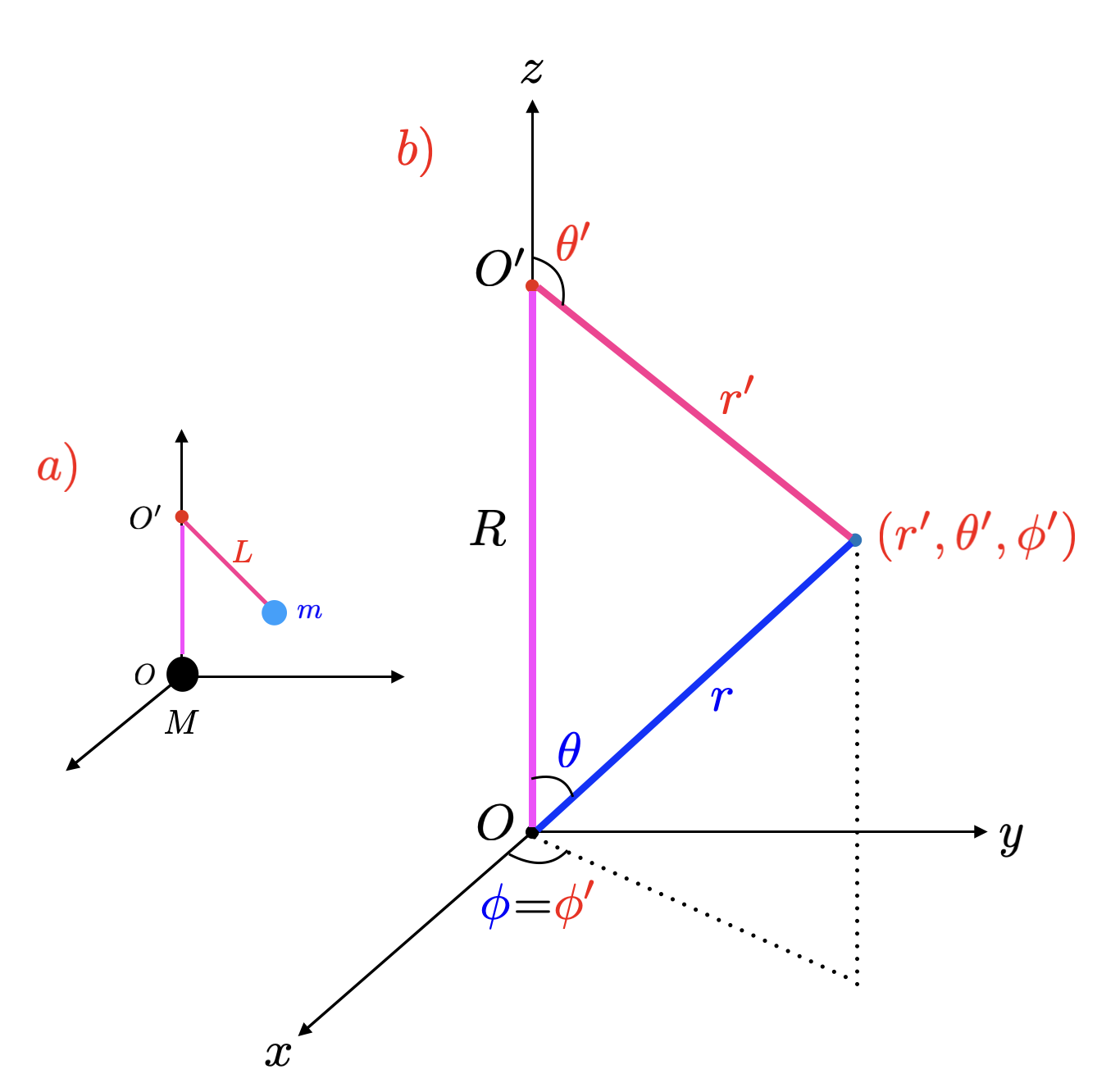}
  \caption{a) Schematic view of a simple pendulum of mass $m$,  length $L$ and support point located at $O'$.   $R$ is the distance from the pendulum to the place $O$ where a mass $M$ is the source of the gravity field in which the pendulum swings. b) Spatial pendulum spherical coordinates $(r', \theta',\phi')$ around $O'$ and its relation to the original spherical coordinates $(r,\theta,\phi)$ about the source mass $M$ corresponding to the Schwarzschild metric \eqref{Schwarzschild} providing the change of coordinates \eqref{pendulum_coordinates}.}
  \label{fig:new_coordinates}
\end{figure}

The simple pendulum (see Fig.\ref{fig:new_coordinates}a)) is one of the simplest examples of a rigid body.  It comprises a mass $m$ tied to a fixed point $0'$ by a massless string of length $L$.
 It is worth mentioning  that the notion of a solid rigid body can be defined in GR by means of the geodesic deviation equation \cite{problem_book} and it has the same number of degrees of freedom as in Newtonian mechanics.   In our case,  we shall denote $\theta'$ its angle deviation measured from the vertical as in Fig.\ref{fig:new_coordinates}b).

Let us consider a source of gravitation corresponding to a mass $M$ considered as a point for simplicity and placed at the origin of coordinates. The solution to the Einstein field equations is the Schwarzschild metric \cite{Carroll, Rindler} written as follows:
\begin{align}\label{Schwarzschild}
c^2 d\tau^2 &= \left (1-\frac{2GM}{c^2 r}\right) c^2 dt^2 - \left(1-\frac{2GM}{c^2 r}\right)^{-1} dr^2\\ \nonumber
                      & - r^2 d\theta^2 - r^2\sin^2\theta d\phi^2,
\end{align}
that is written in spherical coordinates $(r,\theta,\phi)$, $t$ is the coordinate time,  $\tau$ is the proper time and $c, G$ are the light velocity and Newton's gravitational constant,  respectively.

To solve the pendulum equations in GR we may still consider the motion of the pendulum's mass occurring in a plane of motion at constant $\phi$.   By symmetry, we may assume that the metric will have the coordinate $\phi$ cyclic and the same for the time $t$. However, the metric would cease to be diagonal in the coordinates $r$ and $\theta$ because the motion in the plane of the pendulum involves such coordinates.

Furthermore,  it is important to notice that the role of the simple pendulum is not to be a source of gravitational field itself, such as a  planet of mass $M$,   but it is a test particle that we use as a clock. Thus, it is not appropriate to write an energy-momentum tensor containing the particle's mass $m$ for that would give rise to a metric with two field sources, $M$ and $m$, and then the coordinates corresponding to that metric would be those of an additional non-test particle,   thereby not the simple pendulum itself.

\subsection{Pendulum Spherical Coordinates}

The situation we address is that of a simple pendulum which is a test particle in the Schwartzshild metric \eqref{Schwarzschild} with the constraint of being held by a string of length $L$. In this way, the mass $m$ does not back-react on the mass $M$.  In order to impose the pendulum's constraint,  it is convenient to make a transformation of spherical coordinates from those around the mass source $M$ at $O$ to those about the fixed point $O'$ of the pendulum denoted as $r',\theta',\phi'$ as in  Fig.\ref{fig:new_coordinates}b).

Let us denote the change of coordinates by
\begin{equation}
r = F(r^\prime,\theta'), \quad \theta = G(r^\prime,\theta'),
\end{equation}
where the functions $F$ and $G$ are expressed implicilty by the following relations that can be derived from the geometry of Fig.\ref{fig:new_coordinates}
\begin{align}\label{pendulum_coordinates}
r^2 &= R^2 +    r^{\prime 2} + 2R r^\prime \cos\theta'        \\ \nonumber
\tan \theta &= \frac{r^\prime \sin\theta'}{R + r^\prime \cos\theta'} \\ \nonumber
\phi & = \phi'
\end{align}
The Schwarzshild metric \eqref{Schwarzschild} in the new coordinates will be denoted as
\begin{equation}
c^2d\tau^2= g'_{00}c^2dt^2 + g'_{11}dr^{\prime 2} + 2g'_{12} dr' d\theta' + g'_{22} d\theta^{\prime 2}.
\end{equation}
It is not diagonal and we have removed the coordinate $\phi'$ since the pendulum movement's is planar.
The value of $r'$ cannot be fixed during the coordinate transformation but later on in the equations of motion.

The components of the Schwarzschild metric tensor in the pendulum spherical coordinates read as follows:
\begin{align}\label{metric}
g'_{00} & := S(r',\theta') = 1-\frac{R_S}{\sqrt{R^2+r^{\prime} 2 + 2Rr^\prime \cos\theta'}}, \\  \nonumber
g'_{11} &=-\frac{S^{-1}(r',\theta')\left( r' + R\cos\theta' \right)^2 + R^2\sin^2\theta'}{R^2+r^{\prime 2} + 2Rr^\prime \cos\theta'}, \\ \nonumber
g'_{22} &= -\frac{S^{-1}(r',\theta') R^2r^{\prime 2} \sin^2\theta' + r^{\prime 2} \left( r' + R\cos\theta' \right)^2}{R^2+r^{\prime 2} + 2Rr^\prime \cos\theta'},\\ \nonumber
g'_{12} &= -\left[\frac{1-S^{-1}(r',\theta') }{R^2+r^{\prime 2} + 2Rr^\prime \cos\theta'}\right] Rr^\prime \sin\theta' \left( r' + R\cos\theta' \right) \\ \nonumber
\end{align}
where $R$ is the coordinate distance of the fixed point of the pendulum $O'$ to the position of the source gravity field $M$ at the origin $O$ before the transformation and $R_S:=\frac{2GM}{c^2}$ is the Schwarzshild radius.  Note that this is a passive coordinate change: the gravitational field source $M$ has not changed location, what we change is the labeling of the coordinates with respect to the new spherical coordinate system centered at the fulcrum of the simple pendulum. If we take $r'=0$ we can see that there is no singularity there. We have changed the coordinates, but not the geometrical arrangement of the objects.

\subsection{The Pendulum Equations of Motion in the Schwarzschild Metric}


When deriving the equations of motion for the pendulum in the new coordinates we must impose the constraint that the length of the pendulum is fixed by $r'=L$. Thus, all proper time derivatives of $r'$ disappear from the geodesic equations, as well as those of $\phi$, which we can also fix to a constant value. Then,  the structure of the equations of geodesic motion is as follows:
\begin{align}
c\ddot{t} + \Gamma^{\prime 0}_{00} c^2 \dot{t}^2 + 2\Gamma^{\prime 0}_{02} c\dot{t} \dot{\theta'} + \Gamma^{\prime 0}_{22}\dot{\theta'}^2 &= 0, \\ \nonumber
\ddot{\theta'} + \Gamma^{\prime 2}_{00} c^2 \dot{t}^2 + 2\Gamma^{\prime 2}_{02} c\dot{t} \dot{\theta'} + \Gamma^{\prime 2}_{22}\dot{\theta'}^2 &= 0, \\ \nonumber
\end{align}
where the dot denotes derivative with respect to the proper time $\tau$ in equation \eqref{Schwarzschild}.
Now using the non-vanishing Christoffel symbols in the pendulum spherical coordinates (see Appendix \ref{Christoffel}), they can be further simplified 
\begin{align}\label{evolution}
c\ddot{t} +  2\Gamma^{\prime 0}_{02} c\dot{t} \dot{\theta'}  &= 0, \\ \nonumber
\ddot{\theta'} + \Gamma^{\prime 2}_{00} c^2 \dot{t}^2 + \Gamma^{\prime 2}_{22}\dot{\theta'}^2 &= 0,.\\ \nonumber
\end{align}
To solve for these equations, we can use the conservation of energy $E$ per unit mass ($m=1$)
\begin{equation}\label{E}
g'_{00} c^2 \dot{t} = E,
\end{equation}
and the expression of the tangent vector (4-velocity) normalized in the new coordinates,
\begin{equation}\label{velocity}
g'_{00} c^2 \dot{t}^2 + g'_{22} \dot{\theta'}^2 = c^2.
\end{equation}
The Newtonian limit of these equations of motion will play a fundamental role in reproducing the well-known equations for the simple pendulum.   
This  limit corresponds to the following conditions:
\begin{equation}\label{Newtonian-Limit}
E/m \rightarrow c^2, \quad R_S \rightarrow 0,
\end{equation}
keeping only the mass contribution to the energy and at the same time,  disregarding the Schwarzshild radius $R_S$ versus any other distance.
\begin{figure*}[t]
  \includegraphics[width=0.45\textwidth]{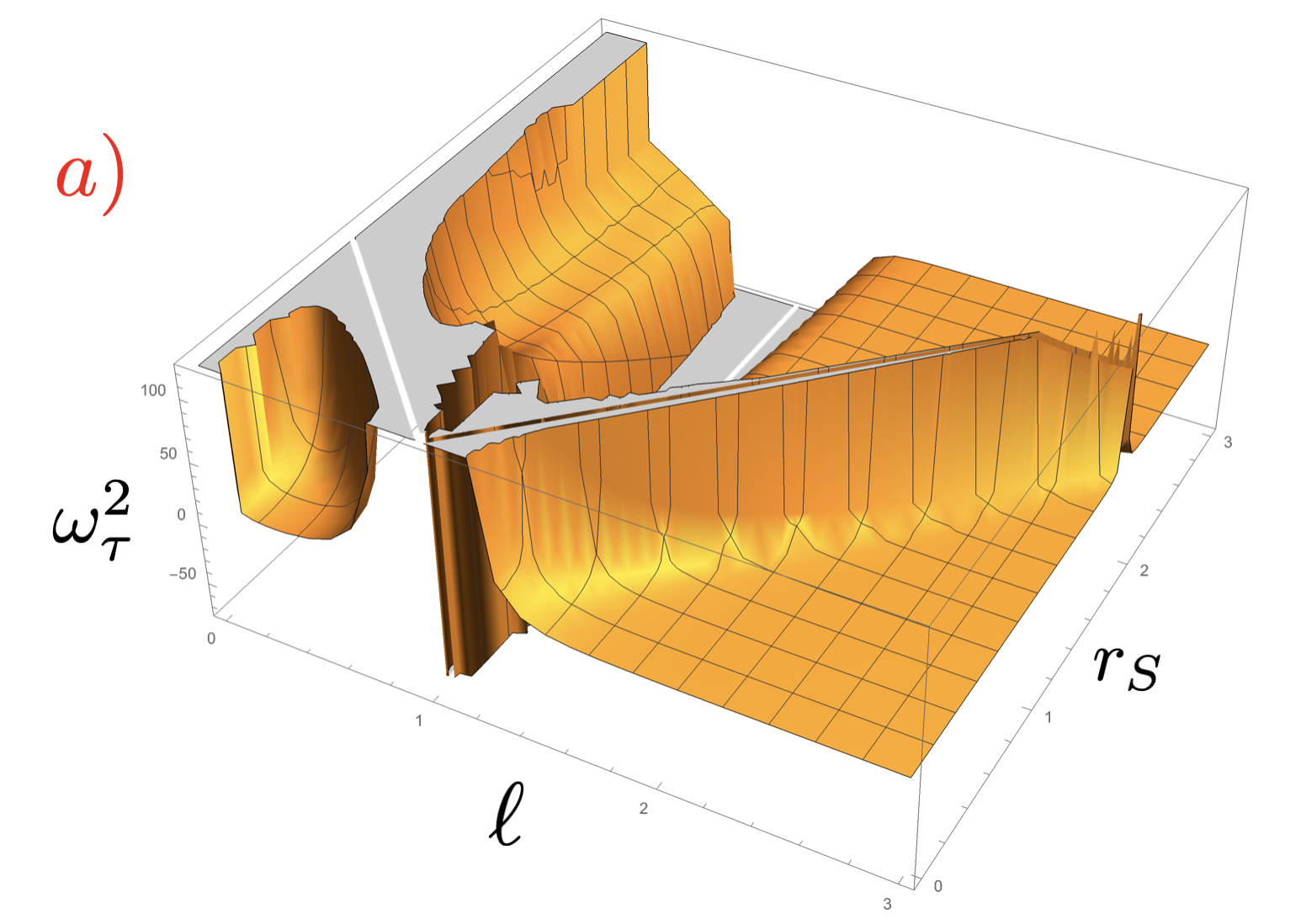}
  \includegraphics[width=0.45\textwidth]{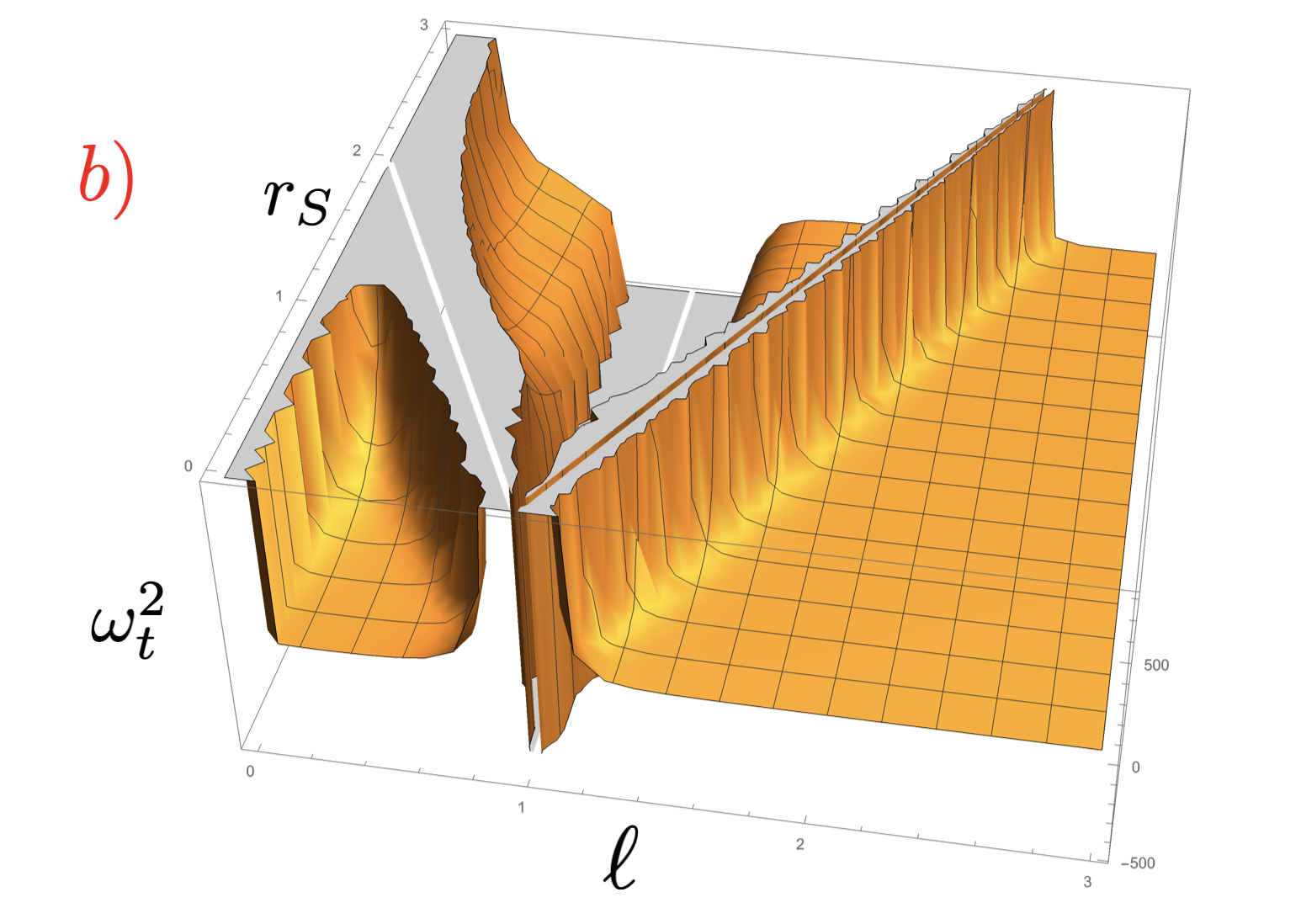}
  \caption{a) Graphic representation of the proper oscillation frequency $\omega^2_\tau$ as a function of the coordinate length $\ell $ and Scharzschild radius $r_S$  in \eqref{frequency2_bis}  measured with respect to the coordinate distance $R$ to the mass $M$ \eqref{reduced_variables}.  The GR divergences in \eqref{frequency2_bis} at reduced parameters  $\ell =1$ and $|\ell -1| = r_S$ \eqref{divergence_tau} are clearly visible.  b) Similar to a) but for the coordinate frequency $\omega^2_t$ \eqref{frequencies} with divergences at $\ell =1$ and $|\ell -1| = \half r_S$ \eqref{divergence_t}.  Notice that in both cases the plots show negative values that are unphysical.  However,  it is proved that they occur when either the support point of the pendulum or its mass or both remain inside the Event Horizon \eqref{negative_values}.}
  \label{fig:landscape_tau}
\end{figure*}
\section{Time Coordinate Geodesic Equation}
\label{sec:time_coordinate}

\subsection{Harmonic Approximation Consistency}

The geodesic equation for the time coordinate in \eqref{evolution} can be solved directly as shown below.
Before that, as we are interested in the isochrony of the simple pendulum in GR,  it is important to check that property in the equation of motion for the time coordinate $t$. Using the constraints from the conservation of energy \eqref{E} and the normalization of the 4-velocity \eqref{velocity} we arrive at:
\begin{equation}\label{t-evolution}
\ddot{t} + 2 \Gamma^{\prime 0}_{02} \frac{E}{g'_{00} c} \sqrt{\frac{g'_{00} c^4 - E^2}{g'_{00} g'_{22} c^4}}=0.
\end{equation}
Upon substitution of the metric components \eqref{metric}, the Christoffel symbol \eqref{0-Christoffel} in \eqref{t-evolution} and linearizing about the equilibrum position $\theta'=\pi$ the amplitud takes the form
\begin{equation}
\ddot{t}  + \theta' \frac{E R R_S}{2c (|R-L|-R_S)^3} \sqrt{\frac{E^2|R-L|}{c^4 (|R-L|-R_S)} - 1} =0.
\end{equation}
Taking the Newtonian approximation \eqref{Newtonian-Limit} in this expression,  the $\theta'$ contribution vanishes due to the form of the square root above recovering the proportionality of both times.

Let $\theta'_0$ be the initial position of the pendulum mass being initially at rest.  Then the energy mass receives a gravitatonal contribution of the form:
\begin{equation}\label{energy}
E = mc^2 - \frac{GMm}{\sqrt{R^2+L^2+2RL\cos\theta'_0}},
\end{equation}
since initially there is no kinetic energy.  Furthermore, for small oscillations about the equilibrium position          $\theta'_0\approx \pi$, the energy can be consistently approximated as
\begin{equation}
\frac{E}{m} \approx c^2 - \frac{GM}{|R-L|},
\end{equation}
It is convenient to express the energy $E$ per unit mass in terms of the relevant lengths of the problem,  namely
\begin{equation}\label{energy_2}
\frac{E}{c^2} \approx 1 - \frac{R_S}{2|R-L|}.
\end{equation}

\subsection{Proper and Coordinate Frequency Relationship}

We can obtain the time coordinate $t$ by integrating the energy conservation condition \eqref{E}
\begin{equation}
t = \frac{E}{c^2} \int \frac{d\tau}{g'_{00}}.
\end{equation}
Within the small oscillation approximation about $\theta'=\pi$,  the metric component $g'_{00}$ \eqref{metric}  can be approximated by the following constant:
\begin{equation}
g'_{00} \approx 1-\frac{R_S}{|R-L|}.
\end{equation}
Using the expression for the energy in the harmonic approximation \eqref{energy_2},  this implies that the time coordinate $t$ and the proper time $\tau$ are proportional one another, namely,
\begin{equation}\label{times}
t = \half \frac{2|R-L|-R_S}{|R-L|-R_S} \tau .
\end{equation}
In the Newtonian limit \eqref{Newtonian-Limit} we recover the identity of times.
Notice that for an external observer, the time diverges when $|R-L|$  approaches $R_S$.

In order to compare oscillation frequencies as measured by the proper time of the pendulum $\tau$,  denoted 
$\omega_\tau$,  with frequencies measured by the coordinate time $t$ of an external observer,  denoted 
$\omega_t$,  we need to express the harmonic approximation of \eqref{evolution}
 in terms of $t$.  This is accomplished by a change of variables expressed as 
\begin{equation}
\ddot{\theta }^{\prime} =( \frac{dt}{d\tau})^2 \frac{d^2}{dt^2} \theta',
\end{equation}
which leads to the harmonic equation in $t$
\begin{equation}
\frac{d^2}{dt^2} \theta' + ( \frac{d\tau}{dt})^2  \omega^2_\tau \theta' = 0.
\end{equation}
Thus,  using \eqref{times} the relation between the proper frequency $\omega_\tau$ and the time-coordinate frequency $\omega_t$ is given by
\begin{equation}\label{frequencies}
\omega_t = 2\left| \frac{|R-L|-R_S}{2|R-L|-R_S}\right| \omega_\tau .
\end{equation}
This gives us the relation between the proper period $T_\tau$ of the pendulum and its coordinate period $T_t$.

\section{Harmonic Oscillations in the Angular Variable}

Now is time to work on the angular equation \eqref{evolution} since we are interested in finding whether it can host an isochronous motion in the limit of small oscillations about the equilibrium point $\theta'=\pi$ in the pendulum spherical coordinates.
This means that we have to get rid off the first derivatives in $\theta'$ and $t$ by using the constraints of
the conservation of energy \eqref{E} and the normalization of the 4-velocity in the new coordinates \eqref{velocity} as we did with the time-coordinate geodesic equation.  Thus,  the equation of motion for the angular variable $\theta'$ of the pendulum in GR turns out to be:
\begin{equation}\label{GR_pendulum}
\ddot{\theta'} + \left[  \frac{\Gamma^{\prime 2}_{00}}{g^{\prime 2}_{00}} - \frac{\Gamma^{\prime 2}_{22}}{g'_{00}g'_{22}}   \right] \frac{E^2}{c^2}  
+ \frac{\Gamma^{\prime 2}_{22}}{g'_{22}} c^2= 0.
\end{equation}

The linearization of this equation about the equilibrium point $\theta'=\pi$ yields the value of the proper frequency $\omega_\tau$ of the pendulum oscillations in General Relativity,
\begin{widetext}
\begin{align}\label{frequency2}
\omega^2_\tau = R_S R\frac{E^2\left[ 2R |R-L| + L\left( |R-L| -R_S \right)  \right]-  2c^4 R \left( |R-L| - R_S \right) }{2c^2 L^2 (R-L)^2 \left( |R-L| -R_S\right)^2 }.
\end{align}
\end{widetext}
Using the conversion factor \eqref{frequencies} we also obtain the expression for the time-coordinate frequency $\omega_t$ for the pendulum.
\begin{figure}[t]
  \includegraphics[width=0.5\textwidth]{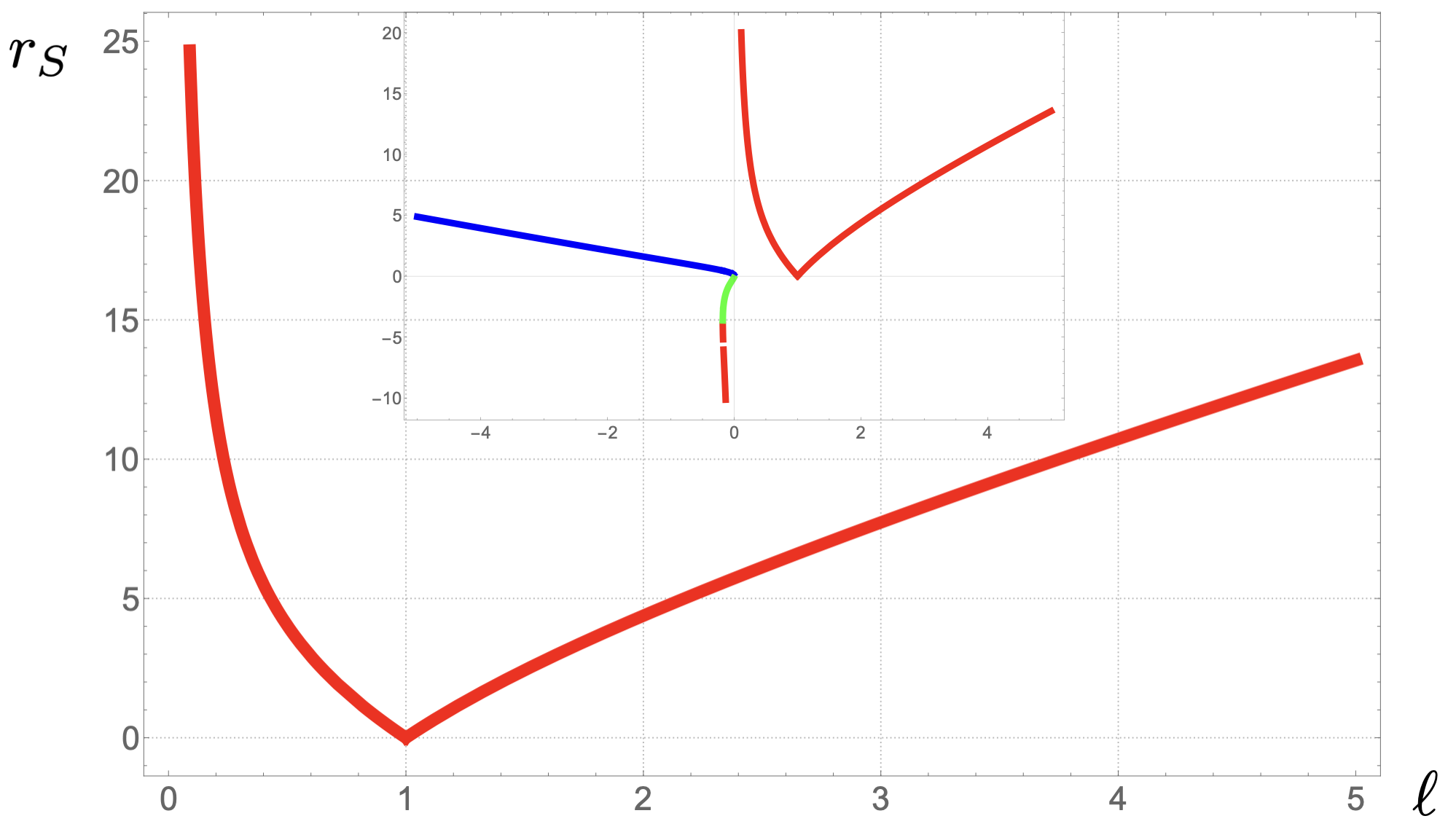}
  \caption{Plot of the zeros of the oscillation frequencies $\omega^2_\tau$ and $\omega^2_t$ as given by the cubic equation \eqref{cubic} for positive values of $\ell $ (reduced length of the pendulum) and $r_S$ (reduced Schwarzschild radius).  This gives rise to the qualitative picture shown in Fig. \ref{fig:zero_pictorial_proper}.  The inset shows the three solutions of the cubic for arbitrary values of the parameters.}
  \label{fig:roots}
\end{figure}
Having a well-defined linearization is not enough to guarantee the isochronous motion of the pendulum.  Additionally,  the frequency must not depend on the initial conditions such as the initial amplitude of the mass $m$.  Apparently, the expression \eqref{frequency2} seems to violate that condition since it depends on the particle's energy $E$, which in turn is dependent on the initial conditions.  However,  as can be seen from the expressions of the energy \eqref{E} and its approximation for small oscillations \eqref{energy_2},  neither the
proper frequency of oscillations nor the coordinate-time frequency does depend on the initial condition.
Then, GR does not break the isochrony and the pendulum can be a good mechanical clock as a matter
of principle.

To test the consistency of the GR expression for the frequency of oscillation \eqref{frequency2}, it must reproduce the corresponding value in Newtonian gravitation for the field generated by  a point source of mass $M$ and also the traditional value of Galileo-Huygens for a constant gravitational field. 
This  limit corresponds to disregarding the gravitational part in the energy \eqref{energy} keeping just the mass contribution and at the same time,  disregarding the Schwarzshild radius $R_S$ versus any other distance.
Taking first the limit in the energy we arrive at the expression,
\begin{align}
\omega^2 \approx GM  \frac{ R \left[  L \left( |R-L| -R_S\right) + 2RR_S  \right] }{ L^2 (R-L)^2 \left( |R-L| -R_S\right)^2 }
\end{align}
and then the limit in the Schwarzshild radius, we finally obtain
\begin{align}\label{frequency_Newtonian}
\omega^2 \approx GM  \frac{ R  }{ L (|R-L|)^3 }.
\end{align}
This is in full agreement with the frequency obtained in Newtonian gravity for the small oscillations associated to the Lagrangian 
\begin{equation}
L/m= \half L\dot{\theta'}^2 + \frac{GM}{\sqrt{R^2+L^2+2RL\cos\theta'}}.
\end{equation}
Furthermore,  when the length of the pendulum is much smaller than the distance to the gravitational source $M$, i.e., $L\ll R$  then
\begin{align}
\omega^2 \approx  \frac{ g }{ L},
\end{align}
with $g=\frac{GM}{R^2}$ the acceleration of gravity.  This is the celebrated Huygens formula for the period of the simple pendulum: $T=2\pi \sqrt{\frac{L}{g}}$.

\section{Anomalies of the Pendulum Clock in GR}
\label{sec:anomalies}

Although we have seen that the pendulum's isochrony is not violated in GR,  there is still the important issue of how accurate can the pendulum be
when working as a mechanical clock for time keeping.  With the first pendulum clock, Huygens was able to measure time with an accuracy of 1 $s$ by choosing
the length of the pendulum of about 1 $m$.  Thus, the accuracy of a given clock increases with its oscillation frequency.  We can use the expression \eqref{frequency2}
to study the accuracy of the pendulum clock depending on the external parameters that govern this behaviour.  
Substituting \eqref{energy_2} into \eqref{frequency2} and simplifying yields,
\begin{widetext}
\begin{align}\label{frequency2_bis}
\omega^2_\tau = GM R\frac{\left(1 - \frac{R_S}{2|R-L|}\right)^2\left[ 2R |R-L| + L\left( |R-L| -R_S \right)  \right]-  2R \left( |R-L| - R_S \right) }{ L^2 (R-L)^2 \left( |R-L| -R_S\right)^2 }
\end{align}
\end{widetext}
The dependency on the Schwarzshild radius $R_S$ represents the novel General Relativistic effect with respect to the Newtonian gravitation \eqref{frequency_Newtonian}.
Next, we study the limiting cases for the  pendulum's frequency as a function of the three relevant coordinate parameters controlling its behaviour: $L,R$ and $R_S$.  Varying $R_S$ amounts to varying the mass $M$ of the gravity source field.  As a general remark,  whenever the GR effects are relevant, we shall avoid to use the notion of proper lengths associated with these parameters since they have originated from a particular coordinate system (Schwarzschild) and in GR coordinates are just labels for events,  not proper times or lengths.  The latter are calculated by means of the metric.  Since there are several possible choices of reference frames or observers,  we shall keep the discussion of the dependency of \eqref{frequency2_bis} on its parameters as simple as possible by considering its direct and plain dependence on them.  In order to obtain a more realistic and meaningful notion of distances we have to choose a certain observer depending on our particular interest.  This could be a free-falling local observer or a shell local observer standing on a shell of a given radius (similar to an observer living on the Earth's surface) and so on and so forth.  Thus,  what we are considering  is a Schwarzschild observer who is just a bookkeeper.
The ultimate meaning of the forthcoming analysis will depend on its conversion to particular local observers who make measurements and convert them to Schwarzschild coordinates for the bookkeeper.
It is in this sense that a general view of the  proper and coordinate frequency landscapes is shown in Fig.\ref{fig:landscape_tau}.

\begin{figure}[t]
  \includegraphics[width=0.5\textwidth]{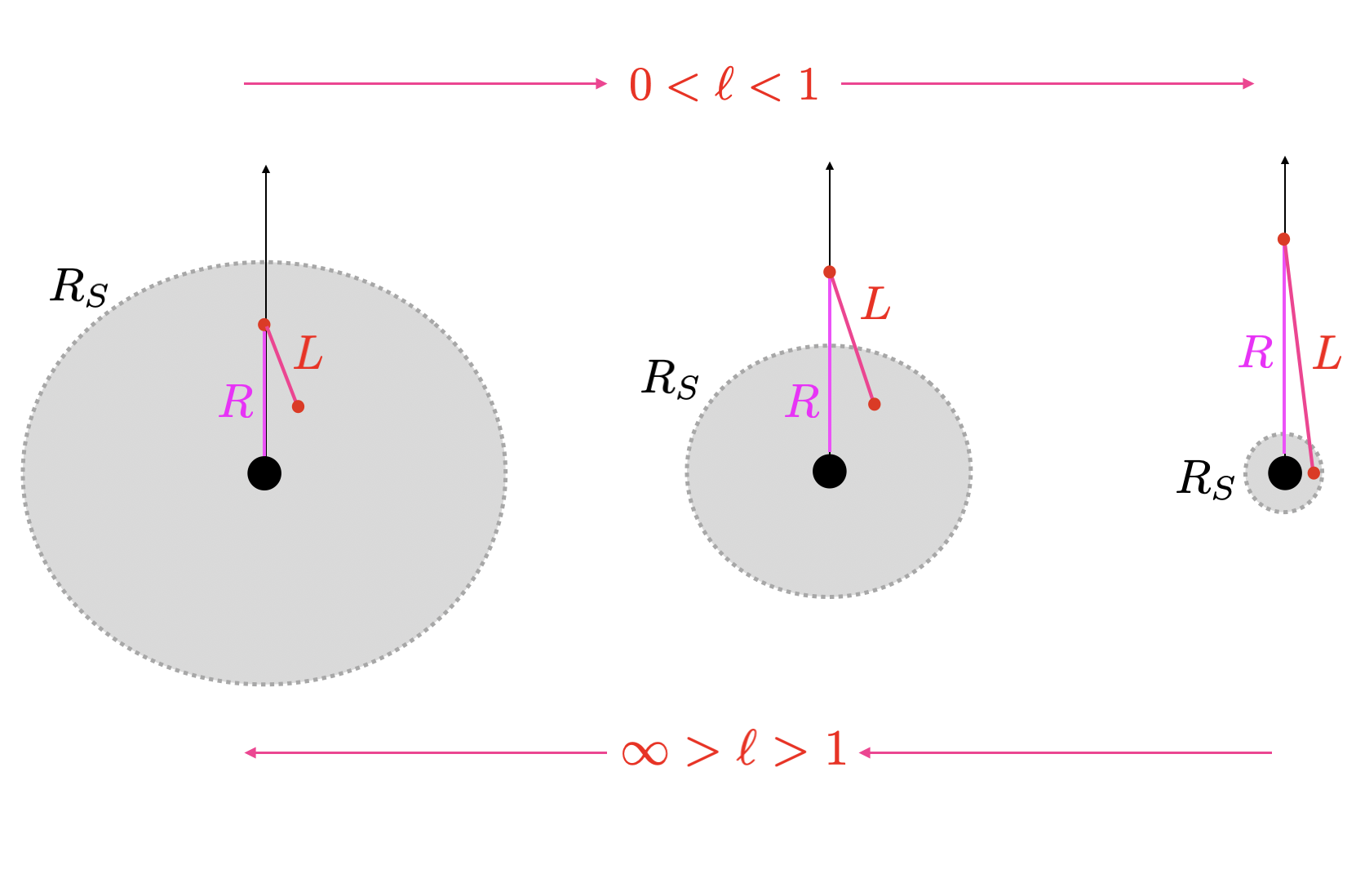}
  \caption{Pictorial representation of the relative relations among the length of the pendulum $L$,   the distance $R$ to the mass $M$ and the Schwarzschild radius $R_S$ when the zeros plotted in Fig.\ref{fig:roots} occur signaling the vanishing of the oscillation frequencies of the pendulum.  This always occurs when some component of the pendulum is inside the Event Horizon.}
  \label{fig:zero_pictorial_proper}.
\end{figure}

\subsection{Diverging Proper  and Coordinate Frequencies}

In the Newtonian expression for the frequency \eqref{frequency_Newtonian},  it is possible to have a singularity by adjusting the length of the pendulum $L$ to match the 
the distance from the pendulum to the gravitational source $R$. This is no surprise because this represents the highly unrealistic situation where the pendulum mass is at the singularity of the gravitational field.  However,  the GR effects allow us to produce a diverging frequency by adjusting the Schwarzschild radius $R_S$ in the denominator of proper frequency $\omega_\tau$ \eqref{frequency2_bis} to match the condition:
\begin{equation}\label{divergence_tau}
|R-L|  = R_S  \neq 0.
\end{equation}
This corresponds to the pendulum mass $m$ grazing the north or the south pole of the event horizon around the mass $M$. Then we would have a very high accuracy clock swinging so fast that has a vanishing period $T$.  These two diverging branches are clearly visible in Fig.\ref{fig:landscape_tau} along with the Newtonian brahch $L=R$.

Notice that this has an important consequence: by varying the length of the pendulum $L$,  keeping $R$ constant,  it is possible to detect the existence of an event horizon. This is very curious because we are using the pendulum's proper time $\tau$, and not the coordinate time $t$ of an external observer.  Note that, in the case of free-fall motion, it does not detect anything anomalous when crossing the event horizon \cite{Misner,Carroll,Rindler}. With a pendulum, we can detect $R_S$ as a singularity in the proper frequency of oscillation.

Interestingly,  when the proper frequency $\omega_\tau$ diverges at \eqref{divergence_tau}, the coordinate time frequency $\omega_t$ remains finite due to the conversion factor \eqref{times}. However,  a different divergence appears in the coordinate frequency at half the Schwarzschild radius
\begin{equation}\label{divergence_t}
 |R-L|  = \half R_S  \neq 0.
\end{equation}
At this value the energy $E$ of the pendulum vanishes \eqref{energy_2} and the conversion factor \eqref{frequencies} diverges. This corresponds to the pendulum mass $m$ grazing half the north or the south pole of the event horizon around the mass $M$, meaning that the pendulum is well inside the EH.  Correspondly,  the proper frequency does not diverge at the Schwarzschild radii \eqref{divergence_t}.

\subsection{Vanishing of Proper and Coordinate Frequencies}

The possibility of vanishing the proper and coordinate oscillation frequencies \eqref{frequencies},  \eqref{frequency2_bis} of the pendulum is a genuinely General Relativistic effect since in Newtonian gravity it cannot occur. This corresponds to a clock that has absolutely lost its accuracy with a period $T$ becoming infinite.  This does not mean that time itself has become frozen but the ability to measure it with a mechanical clock.  This raises the question as to what extent can General Relativity be considered an intrinsically complete theory in cases in which even though isochrony is respected by the mechanical clock,  its poor accuracy makes it useless for time keeping.  Because of the frequency conversion factor \eqref{frequencies},  both the proper and coordinate oscillation frequencies share the same vanishing values in \eqref{frequency2_bis}.

To gain more intuition about this possibility, let us search for these situations by finding the zeros of \eqref{frequency2_bis} whenever $|R-L|\neq 0$ and $R_S\neq |R-L|, 2  |R-L|$ avoiding the singularities.

We shall consider the Schwarzschild radius $R_S$ as the main variable since it represents the novel GR effects.  The vanishing condition of the frequency \eqref{frequency2_bis} yields the following  equation in $R_S$:
\begin{widetext}
\begin{equation}
\left(1 - \frac{R_S}{2|R-L|}\right)^2\left[ 2R |R-L| + L\left( |R-L| -R_S \right)  \right]-  2R \left( |R-L| - R_S \right) =0.
\end{equation}
\end{widetext}
This becomes a cubic equation in $R_S$.  In the following,  it is convenient to deal with dimensionless variables by measuring $R_S$ and $L$ in units of $R$.
Thus, introducing the new variables 
\begin{equation}\label{reduced_variables}
r_S:= \frac{R_S}{R}, \quad \ell := \frac{L}{R},
\end{equation}
we can write the cubic equation as
\begin{equation}\label{cubic}
\frac{\ell}{4(1-\ell )^2} r^3_S - \frac{5\ell +2}{4|1-\ell|} r^2_S +  2\ell r_S  
- |1-\ell | =0.
\end{equation}
Notice that the coefficients of this cubic cannot vanish for $\ell >0$ and $\ell \neq 1$.  The solution of this cubic equation must also fulfill the condition $0<r_S\neq |1-\ell|, \half |1-\ell|$.  The realistic limit case $\ell \ll 1$ does not provide real solutions.  Furthermore,  we may seek approximate solutions by demanding $r_S\ll \half |1-\ell |$ in \eqref{frequency2_bis}.  Then the vanishing condition turns out to be
\begin{equation}
r_S \approx |1-\ell| (\frac{\ell}{\ell -2}) \ll \half |1-\ell |.
\end{equation}
However, these solutions are unphysical since the reduced length $\ell $ becomes negative.

In order to obtain exact solutions to the cubic equation of the Schwarzschild radii \eqref{cubic} we use numerical methods to plot its solutions and search for results fulfilling all the constraints.  The three roots  are shown in Fig.\ref{fig:roots} for positive values in the positive quadrant.  Notice that the realistic values $r_S$ and $\ell $ correspond to the small region around the origin that are values much smaller than the distance of the support point of the pendulum to the field source $M$. 

As the length parameter is close to $\ell \approx 0$ then the reduced Schwarzschild radius $r_S  \rightarrow \infty $.  When $\ell $ starts growing to 1,  $r_S$ starts decreasing.  At a value of 
$\ell = 0.838662$ the reduced Schwarzschild radius $r_S =1$ reaches the fulcrum of the pendulum.  Then,  for 
$0.838662 < \ell < 1$, the  Schwarzschild radius reduces to 0 but it always encloses the mass $m$ of the pendulum as shown in Fig.\ref{fig:zero_pictorial_proper}. Conversely,  as the parameter length of the pendulum starts becoming larger than the parameter distance of the fulcrum to the source mass $M$,  namely $\ell >1$,  the Schwarzschild radius starts growing up rapidly and when $\ell = 1.18792$,  it reaches again $r_S =1$.  From that onwards,  the Schwarzschild radius is always bigger than $\ell -1$, meaning that the fulcrum of the pendulum is always trapped inside the Event Horizon.

Another source of instabilities could arise in the oscillation frequencies if the values of $\omega^2_\tau$ in \eqref{frequency2_bis} become negative ruining the harmonic oscillator behaviour.  In fact,  from Fig.\ref{fig:landscape_tau} it is clearly visible the existence of those negative values.  For example,  when $\ell =0.9$ and $r_S=0.7$ we obtain $\omega^2_\tau = -317.215$ (in reduced units $GM=R=1$).  The analysis of situations where this happens is rather cumbersome given the intricate structure of the oscillation frequencies.  However,  with the help of the Mathematica programming language is possible to show that the set of inequalities,
\begin{equation}\label{negative_values}
\omega^2_\tau <0, \ \ell , r_S >0, \ r_S\neq |\ell -1|, \half |\ell -1|, \  r_S<|\ell -1|,
\end{equation}
yield no solutions.  This means that the instabilities  in the proper and coordinate frequencies only occur under extreme conditions: either the fulcrum or the pendulum mass or both remain inside the Event Horizon.

\section{Conclusions}
\label{sec:conclusions}

Galileo is credited to be the first one to realize that isochrony was the crucial property that allowed
a simple pendulum to be used as a mechanical clock.  Subsequently,  Huygens constructed the first pendulum clock and revolutionized the part of metrology devoted to time keeping.

One of the distinctive features of the theory of General Relativity (GR) is the necessity
to abandon time as an absolute magnitude independent of the reference frame.  In turn, this translates
into the need to specify which system of reference or observer we are using to specify time.  What is always
assumed in GR is that we have a clock at our disposal to measure time in any circumstance.  Here we have asked the question of what is the fate of isochrony of the simple pendulum in the hypothetical circumstances in which the pendulum is subject to truly GR effects.  This study is ideal in many regards.  For instance, the string of the pendulum is assumed to be strong enough to keep up with the gravitational pull without breaking.

We have found that  in GR we can use a simple pendulum as a mechanical clock abiding by its laws and, as a matter of principle,  to measure time within the theory itself  without resorting to other devices whose functioning is outside the theory of GR,  like clocks based on quantum mechanics.  It is in this sense that GR can be considered a complete theory with respect to measuring time.  In doing so,  this study allows us to make a tour through the history of classical mechanics from Einstein to Newton and Galileo.
 
We have provided explicit formulas for the proper and coordinate oscillation frequencies of the simple pendulum subject to the gravity field of a point-like mass $M$ as a function of its coordinate parameters $L$ (length of the pendulum),  $R$ (distance to the gravity source $M$) and the Schwarzschild radius $R_S$  \eqref{frequency2},\eqref{frequency2_bis}.    Thus,  several unstable situations are identified that may yield divergences,  zeros or imaginary values for the oscillation frequencies  \eqref{divergence_tau},\eqref{divergence_t}, \eqref{cubic}, \eqref{negative_values}.  
 In particular,  the set of Schwarzschild radii that may produce vanishing values of the oscillation frequencies satisfy a cubic equation whose coefficients depend on the length $L$ and distance $R$ \eqref{cubic}.
 Interestingly,  all these anomalies get hidden inside the Event Horizon.

The existence of isochronous motion compatible with GR can be posed on general grounds by means of equations \eqref{harmonic2} and \eqref{geodesics}.This notion of isochrony is strong in the sense that no approximations are made,  like small oscillations about an equilibrium position.  Thus,  it remains open to find solutions of the Einstein field equations breaking isochrony with purely gravitational interactions.
In addition to the simple pendulum, there are other alternatives of mechanical devices to make a clock in GR like the torsional pendulum but the external force is not gravitational in origin.

We have followed a more practical approach and seeked solutions to the geodesic equations that can host isochronous motion as well in the case of small oscillations.  Our methodology,  being analytical,  relies heavily on symmetries of the metric tensor such as time $t$ independence  (energy)  and $\phi $ independence (angular momentum).  With eqs.  \eqref{pendulum_coordinates} it is also possible to study GR effects in the spherical pendulum,  in which the angle $\phi$ becomes dynamical.  It is conceivable that this method can be applied to other instances of metrics like the Reissner-Nordstrom and Kerr-Newman.  The latter would be the way to find the generalization of the Foucault pendulum in GR in the chargeless case.  Similarly,  a mathematically interesting problem is to check whether the non-linear function in the harmonic approximation to the angular equation of motion \eqref{GR_pendulum} can be expressed in terms of elliptic functions since it is constructed by means of trigonometric functions but in a rather intricate way.

\acknowledgements
We acknowledge support from the CAM/FEDER Project No.S2018/TCS-4342 (QUITEMAD-CM), Spanish MINECO grants MINECO/FEDER Projects,  PGC2018-099169-B-I00 FIS2018, 
MCIN with funding from European Union NextGenerationEU (PRTR-C17.I1) an Ministry of Economic Affairs Quantum ENIA project.  M. A. M.-D. has been partially supported by the U.S.Army Research Office through Grant No. W911NF-14-1-0103.

\appendix

\section{Schwarzschild Metric in the Pendulum Coordinates}

To derive the new components of the Schwarzschild metric in the pendulum spherical coordinates $(ct,r',\theta',\phi)$, let us perform the change of coordiantes from eqs. \eqref{pendulum_coordinates}
\begin{align}
g'_{00} &= S(r',\theta'), \\  \nonumber
g'_{11} &=- \left[ S^{-1}(r',\theta') F^2_{r'}(r',\theta') + F^2(r',\theta')  G^2_{r'}(r',\theta')\right], \\  \nonumber
g'_{22} &=- \left[ S^{-1}(r',\theta') F^2_{\theta'}(r',\theta') + F^2(r',\theta')  G^2_{\theta'}(r',\theta')\right], \\  \nonumber
g'_{12} &=- [ S^{-1}(r',\theta') F_{r'}(r',\theta')F_{\theta'}(r',\theta') \\  \nonumber
             &+ F^2(r',\theta')  G_{r'}(r',\theta')G_{\theta'}(r',\theta')],
\end{align}
where the subscripts $r'$ and $\theta'$ in functions $F,  G$ denote the partial derivatives with respect to those coordinates. 
Using the pendulum spherical coordinates \eqref{pendulum_coordinates},  the expressions of the functions $F,G$ and their derivatives take the following form:
\begin{align}
F^2(r',\theta') &= R^2+r^{\prime 2} + 2Rr^\prime \cos\theta' \\ \nonumber
F^2_{r'}(r',\theta') &= \frac{\left( r' + R\cos\theta' \right)^2}{R^2+r^{\prime 2} + 2Rr^\prime \cos\theta'} \\ \nonumber
F^2_{\theta'}(r',\theta') &= \frac{\left( R  r'\sin\theta' \right)^2}{R^2+r^{\prime 2} + 2Rr^\prime \cos\theta'} \\ \nonumber
F_{r'}(r',\theta')F_{\theta'}(r',\theta') &= - \frac{ \left( r' + R\cos\theta' \right) }{R^2+r^{\prime 2} + 2Rr^\prime \cos\theta'} R  r'\sin\theta' \\ \nonumber
G_{r'}(r',\theta') &= \frac{R  r'\sin\theta' }{R^2+r^{\prime 2} + 2Rr^\prime \cos\theta'} \\ \nonumber
G_{\theta'}(r',\theta') &=  \frac{r' \left( r' + R\cos\theta' \right) }{R^2+r^{\prime 2} + 2Rr^\prime \cos\theta'} \\ \nonumber
\end{align}

\section{New Christoffel Symbols}
\label{sec:Christoffel}

We need to compute the Christoffel symbols for the Schwarzschild metric in the pendulum spherical coordinates.  As now the metric is no longer diagonal, this amounts to computing first the inverse of the following $2\times 2$ matrix:
\begin{equation}
g'_{2\times 2} := \begin{pmatrix}
g'_{11} & g'_{12} \\
 g'_{12} &  g'_{22},
\end{pmatrix}
\end{equation}
namely,
\begin{equation}
(g'_{2\times 2})^{-1} := \frac{1}{g'_{11}g'_{22} - (g'_{12})^2}\begin{pmatrix}
g'_{22} & -g'_{12} \\
 -g'_{12} &  g'_{11},
\end{pmatrix}
\end{equation}
Then, the Christoffel symbols in the new coordinates are computed starting from its definition:
\begin{equation}\label{Christoffel}
\Gamma^{'\lambda}_{\mu \nu} := \frac{1}{2} g^{\prime \lambda \alpha} \left( \frac{\partial g'_{\alpha \mu }}{\partial x^{\prime \nu }}  +  \frac{\partial g'_{\alpha \nu}}{\partial x^{\prime \mu}}  -  \frac{\partial g'_{\mu \nu}}{\partial x^{\prime \alpha}}\right)
\end{equation}
where $x^{\prime \mu}$ are the pendulum spherical coordinates defined in \eqref{pendulum_coordinates}.

We shall be interested in the Christoffel symbols appearing in the geodesic equations of motion for the new angular coordinate $\theta'$ and the time coordinate $t$, corresponding to the superscripts $\lambda=2,0$ in eq. \eqref{Christoffel}. For the angular coordinate, we find
\begin{align}
\Gamma^{'2}_{00} &=  -\frac{1}{2} \frac{1}{g'_{11}g'_{22} - (g'_{12})^2} \left( -g'_{12} \frac{\partial g'_{00} }{\partial r'} +  g'_{11} \frac{\partial g'_{00} }{\partial \theta'} \right), \\ \nonumber
\Gamma^{'2}_{02} &= 0,   \\ \nonumber
\Gamma^{'2}_{22} &=  \frac{1}{2} \frac{1}{g'_{11}g'_{22} - (g'_{12})^2}\left[- \left( 2   \frac{\partial g'_{12} }{\partial \theta'} -  \frac{\partial g'_{22} }{\partial r'} \right)
+ g'_{11}   \frac{\partial g'_{22} }{\partial \theta'}  \right].  \\ \nonumber
\end{align}
As for the Christoffel symbols for the time coordinate, we have
\begin{align}\label{0-Christoffel}
\Gamma^{'0}_{00} &=  0, \\ \nonumber
\Gamma^{'0}_{22} &= 0,   \\ \nonumber
\Gamma^{'0}_{02} &=  \frac{1}{2 g'_{00}}   \frac{\partial g'_{00} }{\partial \theta'}, \\ \nonumber
\end{align}

\section{Quartics for the Coordinate Length of the Pendulum}
In section \ref{sec:anomalies} we have analysed the vanishing of the oscillation frequencies as a function of the Schwarzschild radii since this is the natural parameter to assess the GR effects in those frequencies through the cubic equation \eqref{cubic}.  Alternatively,  we can also use the coordinate length parameter $L$ to carry out that study.  For completeness,  we write the quartic equation for the zeroes of the oscillation frequencies in terms of this parameter as follows.   For $\ell <1$, we obtain
\begin{widetext}
\begin{equation}
4 \ell^4  - 4 \ell^3 (3 + 2 r_S) + \ell^2 (12 + 16 r_S + 5 r_S^2) - 
 \ell (4 + 8 r_S + 3 r_S^2 + r_S^3) - 2r_ S^2 = 0,
\end{equation}
\end{widetext}
while for $\ell >1$,
\begin{widetext}
\begin{equation}
4 \ell^4  + 4 \ell^3 (-3 + 2 r_S) + \ell^2 (12 - 16 r_S + 5 r_S^2) + 
 \ell (-4 + 8 r_S - 3 r_S^2 +r_ S^3) - 2 r_S^2 =0.
\end{equation}
\end{widetext}


\end{document}